\documentclass[a4paper,11pt]{article}
\usepackage{jheppub} 

\usepackage{graphicx}
\usepackage{amsmath}
\usepackage{amssymb}
\usepackage{comment}
\usepackage{hyperref}

\usepackage{subfigure}
\newcommand{\bi}{\begin{itemize}}
\newcommand{\ei}{\end{itemize}}
\newcommand{\be}{\begin{enumerate}}
\newcommand{\ee}{\end{enumerate}}

\newenvironment{dfn}{{\vspace*{1ex} \noindent \bf Definition }}{\vspace*{1ex}}

\newcommand{\nn}{\nonumber}  %

	\newcommand{\beq}{\begin{eqnarray}}
	\newcommand{\eeq}{\end{eqnarray}}
	\newcommand{\bea}{\begin{eqnarray}\begin{aligned}}
	\newcommand{\eea}{\end{aligned}\end{eqnarray}}

\title{\boldmath From $G_2$ to $SO(8)$: Emergence and reminiscence of supersymmetry and triality}

\author[a]{Zhi-Qiang Gao}\emailAdd{zqgao@berkeley.edu}
\author[b,c,d,e,1]{and Congjun Wu\note{Corresponding author.}}\emailAdd{wucongjun@westlake.edu.cn}
\affiliation[a]{Department of Physics, University of California, Berkeley, California 94720, USA}
\affiliation[b]{New Cornerstone Science Laboratory, Department of Physics, School of Science, Westlake University, Hangzhou 310024, Zhejiang, China}
\affiliation[c]{Institute for Theoretical Sciences, Westlake University, Hangzhou 310024, Zhejiang, China}
\affiliation[d]{Key Laboratory for Quantum Materials of Zhejiang Province, School of Science, Westlake University, Hangzhou 310024, Zhejiang, China}
\affiliation[e]{Institute of Natural Sciences, Westlake Institute for Advanced Study, Hangzhou 310024, Zhejiang, China}
\abstract{
We construct a (1+1)-dimension 
continuum model of 4-component fermions incorporating the exceptional Lie group symmetry $G_2$.
Four gapped and five gapless phases 
are identified via the one-loop renormalization group analysis.
The gapped phases are controlled by four different stable $SO(8)$ Gross-Neveu fixed points, among which three exhibit an emergent triality, while the rest one possesses the self-triality, i.e., invariant under the triality mapping. 
The gapless phases include three $SO(7)$ critical ones, a $G_2$ critical one, and a Luttinger liquid. 
Three $SO(7)$ critical phases correspond to different $SO(7)$ Gross-Neveu fixed points connected by the triality relation similar to the gapped SO(8) case.
The $G_2$ critical phase is controlled by an unstable fixed point described by a direct product of the Ising and tricritical Ising conformal field theories with the central charges $c=\frac{1}{2}$ and $c=\frac{7}{10}$, respectively, while the latter one is known to possess spacetime supersymmetry.
In the lattice realization with a Hubbard-type interaction, the triality is broken into the duality between two $SO(7)$ symmetries and the supersymmetric $G_2$ critical phase exhibits the degeneracy between bosonic and fermionic states, which are reminiscences of the continuum model. 
}
\begin{document}
\maketitle
\flushbottom

\section{Introduction}
\label{sec:intro}

Exceptional Lie groups play an essential role in the study of modern condensed matter physics~\cite{zamolodchikov1989,coldea2010,bernevig2003,
lopes2019}. $E_8$, the largest exceptional Lie group, is employed to describe an exotic quantum Hall state~\cite{Kitaev2006}.
The $E_8$ excitations are also observed in an anti-ferromagnetic Ising spin chain~\cite{coldea2010}. On the other hand, the smallest exceptional Lie group $G_2$ with rank-2 and 14 dimensional~\cite{agricola2008}, which is the automorphism group of the non-associative algebra octonion~\cite{baez2002}, has especially attracted many interests recently~\cite{Hu2018,Gao_2020,Luo2020,Li2023}. 
In particular, a lattice model which has an explicit $G_2$ symmetry based on 4-component, such as spin-$\frac{3}{2}$, fermions is constructed~\cite{Gao_2020}. 
However, $G_2$ symmetry is shown to be intrinsically strongly correlated, that it cannot be realized in non-interacting systems~\cite{Gao_2020}, which makes it hard to study analytically. 

Inspired by Ref.~\cite{rzfa}, we investigate the coarse-grained version of the lattice $G_2$ model~\cite{Gao_2020} in (1+1) spacetime dimension, where many powerful tools can be utilized to study strong correlation systems. 
Through an one-loop renormalization group (RG) calculation, we reveal the hidden $SO(8)$ triality structure. 
The phase diagram is consistent with the mean-field analysis in the lattice model~\cite{Gao_2020}. 
Interestingly, we also find in the phase diagram that one critical phase is described by the tricritical Ising (TCI) conformal field theory (CFT)~\cite{vafa1995}. 
TCI CFT actually has spacetime supersymmetry~\cite{FRIEDAN1985,QIU1986}, which is corresponding to the degeneracy between bosonic and fermionic states in the lattice model~\cite{Gao_2020}. 
Coset constructions based on $G_2$ Wess-Zumino-Witten (WZW) models realizing exotic non-Abelian statistics~\cite{Hu2018,Luo2020,Li2023} can also emerge in our model. 
Moreover, our model also exhibits the exotic phenomena of multiversality and unnecessary criticality~\cite{Bi2019,Abhishodh:2023}.

The rest part of this article is organized as follows:
In Section \ref{sect:model}, the models are introduced both in the form of the lattice Hamiltonian and in the coarse-grained continuum version.
In Section \ref{sect:RG}, the one-loop RG analyses are perform to yield the RG flow patterns and the corresponding gapped and critical phases. 
The embedding of the lattice model in the phase diagram of the continuum model is presented in Section \ref{sect:embedding}.
Conclusions are presented in Section \ref{sect:conclusions}.

\section{The model Hamiltonians}
\label{sect:model}

\subsection{The minimal $G_2$ symmetric lattice model}

In a preceding work~\cite{Gao_2020}, we construct a minimal lattice model with the $G_2$ symmetry based on  4-component spin-$\frac{3}{2}$ fermions as briefly reviewed below. 
The 4-component spinor $\psi_\sigma(i)$ defined on each site $i$ with 
$\sigma=\pm\frac{3}{2},\pm \frac{1}{2}$ spans a local $SO(8)$ algebra $\{N_{ab}(i)\}$ with $0\le a, b\le 7$~\cite{wu2003,wu2006,Gao_2020} . 
Here each generator $N_{ab}(i)$ is a fermion bilinear operator in the particle-hole or particle-particle channel. 

An $SO(5)$, or, isomorphically, $Sp(4)$ subalgebra of $SO(8)$ is simply $L_{ab}=N_{ab}=\frac{1}{2}\psi^\dagger(i) \Gamma_{ab} \psi(i) $ with 
$1\le a, b\le 5$, where $\Gamma_{ab}$'s are the commutators
among the five rank-2 anti-commuting $\Gamma$-matrices, i.e.,  
$\Gamma_{ab}=\frac{i}{2}[\Gamma_a, \Gamma_b]$ with $\{\Gamma_a, \Gamma_b\}=2\delta_{ab}$.
The 10 generators of $Sp(4)$ unify both spin and spin-octupole operators based on the spin-$\frac{3}{2}$
spinor $\psi$.
An $SO(7)$ subalgebra, 
denoted as $SO(7)_A$ below, is an extension of $Sp(4)$, defined as 
$M_{ab}(i)=N_{ab}(i)$ with $0\le a, b\le 6$.
It further includes, and hence, unifies the particle number
$M_{06}(i)=\frac{1}{2}(\psi^\dagger(i) \psi(i) -2)$, and the pairing operators in the spin quintet channel
$M_{0a}(i)=\frac{1}{2}\psi^\dagger(i) \Gamma_a R \psi^\dagger(i)$
with $1\le a\le 5$ where $R$ is the charge conjugation matrix
satisfying $R^\mathbf{T}=R^{-1}=-R$.
Pairing operators are complex, and one complex operator is equivalent to two real ones.
The rest 7 operators denoted $V_a(i)=N_{a7}(i)$ form a vector representation of the $SO(7)_A$ defined above,
unifying 5 spin quadrupole operators with
$N_{a7}=\frac{1}{2}\psi^\dagger(i) \Gamma^a \psi_(i)
$ with $1\le a \le 5$ and 1 complex singlet pairing operator 
$N_{07}=\psi^\dagger(i) R \psi^\dagger (i)$.
For details please refer to appendix~\ref{app:A}. 
  
The $SO(7)_A$ algebra defined above can be decomposed into a 14-dimensional $G_{ab}$ part spanning the $G_2$ subalgebra and a 7-vector $T_a$ belonging to the coset $SO(7)_A/G_2$ defined as~\cite{Gao_2020}
\beq
G_{ab}(i)=\frac{2}{3}  M_{ab}(i) +\frac{1}{6} C_{abcd}
 M_{cd}(i),~~~
T_a(i)=\frac{1}{2\sqrt{3}} C_{abc}  M_{bc}(i),
\eeq
where $C_{abcd}$ is the structure constant of the non-associative octonion algebra. 
Note that $G_{ab}(i)$ satisfy 7 constraints $C_{abc}G_{ab}(i)=0$, where 
$C_{abc}=\frac{1}{24}\epsilon_{abcdefg}C_{defg} $ is the dual tensor of $C_{abcd}$.
Therefore, only 14 of them are linear independent. 
$T_a(i)$ transform under the vector representation of $G_2$~\cite{gunaydin1996,Gao_2020}.
$V_a(i)$ defined before also form a $G_2$ vector representation~\cite{gunaydin1996,Gao_2020}. 
Inversely, the $SO(7)_A$ generators can be represented by $G_{ab}(i)$ and $T_a(i)$ as
\beq
M_{ab}(i)=G_{ab}(i) + \frac{1}{\sqrt 3} C_{abc} T_c(i).
\eeq

Another $SO(7)$ algebra denoted as $SO(7)_B$ is constructed below
sharing the same $G_2$ subalgebra as in $SO(7)_A$.
Define operators $T_a^\prime (i)$ and $V_a^\prime (i)$ as a ``120$^\circ$ rotation" of $T_a(i)$ and $V_a(i)$: 
\beq
T_a^\prime(i)+i V_a^\prime(i)= e^{i\frac{2}{3}\pi} (T_a(i)+iV_a(i)).
\eeq
The generators of $SO(7)_B$ are expressed as~\cite{Gao_2020},
\beq
M_{ab}^\prime(i)=G_{ab}(i) + \frac{1}{\sqrt 3} C_{abc} T_c^\prime(i),
\label{eqMab}
\eeq
and $V_a^\prime (i)$ form an vector representation of $SO(7)_B$. 
The Casimirs of these two $SO(7)$'s are connected by the duality relations~\cite{Gao_2020},
\bea
C_A(i)= \sum_{0\le a<b \le 6} M_{ab}(i) M_{ab} (i) = \frac{1}{4}
\sum_{abcd} C_{abcd} M^\prime_{ab}(i) M^\prime_{cd}(i), \\
C_B(i)= \sum_{0\le a<b \le 6} M^\prime_{ab}(i) M^\prime_{ab} (i) =
\frac{1}{4} \sum_{abcd} C_{abcd} M_{ab}(i)M_{cd}(i).
\eea

A lattice model $H=H_0+H_{int}^{G_2}$ realizing the $G_2$ symmetry 
has been constructed 
as the intersection between $SO(7)_A$ and $SO(7)_B$ symmetries~\cite{Gao_2020},
\bea
&H_0 = -t_\text{hop} \sum_{\langle ij \rangle, \sigma} \Big(\psi^{\dag}_{\sigma}(i)
\psi_{\sigma}(j)+h.c. \Big), \\
&H^{G_2}_\text{{int}}=
u\sum_{i}C_A(i)+v\sum_{i}C_B(i),
\label{eq2}
\eea
where $u$ and $v$ are coupling constants. When $v=0$, or, $u=0$, the lattice model of (\ref{eq2}) restores the $SO(7)_A$, or, $SO(7)_B$ symmetry, respectively. 

Besides the $SO(7)_A$ and $SO(7)_B$ symmetries in the lattice model of (\ref{eq2}),  there exists a third $SO(7)_M$ one sharing the same $G_2$ sub-group, where
$M$ stands for Majorana.
Its generators $M_{ab}^{\prime\prime}(i)$ can be similarly defined through (\ref{eqMab}), with $T_a^{\prime}(i)$ and $V_a^{\prime}$
substituted by $T_a^{\prime\prime}(i)$
and 
$V_a^{\prime\prime}$
defined by ``$-120^\circ$ rotation", i.e., 
$
T_a^{\prime\prime}(i)+i V_a^{\prime\prime}(i)= e^{-i\frac{2}{3}\pi} (T_a(i)+iV_a(i)).
$
The $SO(7)_M$ algebra can be naturally expressed in terms of Majorana fermions constructed based on the 4-component spinor $\psi_\sigma(i)$, i.e.,  
$\chi_m(i)$ with $0\le m\le 7$ (see appendix~\ref{app:A} for a detailed definition).
By straightforward calculations, we express $M_{ab}^{\prime\prime}$ and $V_a^{\prime\prime}(i)$ as
\beq
M_{ab}^{\prime\prime}(i)=i\chi_m(i)\chi_n(i), ~~~
V_a^{\prime\prime}(i) =i\chi_0(i)\chi_m(i),
\eeq
where $1\le m\le 7$.
As $(i\chi_m(i)\chi_n(i))^2$ is always equal to 1, the Casimir of $SO(7)_M$ is a $c$-number. 
Therefore, it is absent in the lattice Hamiltonian (\ref{eq2})  where the on-site Hubbard-type interactions are written in terms of Casimirs. Nevertheless, in the coarse-grained continuum model, the $SO(7)_M$ symmetry will emerge at low energy and fulfill the $SO(8)$ triality relation.

\subsection{The coarse-grained continuum model in (1+1)D}

In (1+1)D, the low energy effective theory of the free Hamiltonian $H_0$ in (\ref{eq2}) is the $SO(8)_1$ WZW model~\cite{Shankar1981} expressed in terms of Majorana fields $\chi_m^{R/L}$ with $0\le m\le 7$. 
These Majorana fields $\chi_m(i)$ defined above form the 8-dimensional vector representation of the $SO(8)$ symmetry. 
The disorder operators of the $SO(8)_1$ WZW model build another two sets of Majorana fermions spanning the 8-dimensional spinor and anti-spinor representations~\cite{Shankar1981}. 
Hence, when the $SO(8)$ symmetry is broken to $SO(7)$, three different choices are possible such that one of the three 8-dimensional representations is decomposed to $8=7\oplus1$ under the $SO(7)$ while the other two remain 8-dimensional.
If this residual $SO(7)$ symmetry is chosen to be $SO(7)_M$, then $\chi_0$ is the singlet invariant under $SO(7)_M$, and $\chi_{1,2,...,7}$ form the 7-dimensional vector of $SO(7)_M$. 
The triality relation among the three 8-dimensional $SO(8)$ representations is inherited by that among $SO(7)$ symmetries.

The chiral currents $G_{ab}^{R/L}$, $T_a^{R/L}$, and $V_a^{R/L}$ in the coarse-grained model are defined as fermion bilinears similar to their lattice cousins, with $\psi_\sigma(i)$ substituted by $\psi_{R,\sigma}(z)$ and $\psi_{L,\sigma}(\bar{z})$ for the right- and left-moving fermions, respectively. 
Note that Umklapp terms $\psi_{R}^\dag \psi_{R}^\dag\psi_L\psi_L$ and $\psi_{L}^\dag\psi_{L}^\dag\psi_R\psi_R$ (spin indices omitted) are allowed, since $H_0$ should be at half-filling to maintain the $G_2$ symmetry. 
Any linear combination of $T_a$ and $V_a$ form a $G_2$ vector. Therefore, in terms of $SO(7)_A$ operators, 
the most general form of the $G_2$ invariant low energy effective Hamiltonian density reads
\beq
\mathcal{H}_\text{int}=g\sum_{a\neq b}G_{ab}^R G_{ab}^L+t\sum_{a}T_a^R T_a^L+y\sum_{a}V_a^R V_a^L+\frac{w}{\sqrt{2}}\sum_{a}
\left(T_a^R V_a^L+V_a^R T_a^L
\right ),\label{eq3}
\eeq
where $g$, $t$, $y$, and $w$ are coupling constants.
The $SO(7)_A$ symmetry appears at 
$w=0$, $t=2g$, and $y$ arbitrary.

To explicitly reveal the triality relation, the Hamiltonian density 
(\ref{eq3}) is rewritten in an equivalent way as
\beq
\mathcal{H}_\text{int}=g\sum_{a\neq b}G_{ab}^R G_{ab}^L+t_A\sum_{a}T_a^R T_a^L +t_B\sum_{a}T_a^{\prime R} T_a^{\prime L}
+t_M\sum_{a}T_a^{\prime\prime R} T_a^{\prime\prime L}, 
\label{eq4}
\eeq
with the relation
\beq
t_A=t-\frac{1}{3}y,~~
t_B=\frac{2}{3}y+\frac{\sqrt{6}}{3}w,~~
t_M=\frac{2}{3}y-\frac{\sqrt{6}}{3}w.
\eeq
The triality of the three $SO(7)$ symmetries is already implied in the permutation of three coupling constants $t_A$, $t_B$, and $t_M$.

\section{Renormalization group flows and fixed planes}
\label{sect:RG}

Based on the $G_2$ current algebra \cite{gunaydin1996},
we perform the one-loop RG calculation in appendix~\ref{app:B},
yielding the following RG equations for the coupling constants defined in (\ref{eq3}),
\bea
&\frac{\text{d}g}{\text{d} l}=16g^2+t^2+y^2+w^2,\\
&\frac{\text{d}t}{\text{d} l}=16gt+2t^2+2y^2-2w^2,\\
&\frac{\text{d}y}{\text{d} l}=16gy+4ty+2w^2,\\
&\frac{\text{d}w}{\text{d} l}=16gw-4tw+4yw.
\label{eqRG1}
\eea
For illustrating of the triality explicitly, the RG equations (\ref{eqRG1}) can be cast in terms of $g$ and $t_i$ with $i \in \{A, B, M\}$ in a more symmetric form as
\bea
&\frac{\text{d}g}{\text{d} l}=16g^2+\sum_i t_i^2+\frac{1}{4}\sum_{i\neq j}t_
it_j, \\
&\frac{\text{d}t_i}{\text{d} l}=16gt_i+2t_i^2+
4  \frac{t_A t_B t_M}{t_i}.
\label{eqRG2}
\eea

An analytical solution to the RG equations (\ref{eqRG2}) would be difficult, however, the geometrical structure of the RG flows is rather clear. 
They exhibit fixed planes (not necessarily the same as critical surfaces) in which the RG flows 
keep in-plane evolutions. 
The fixed planes can be classified into two types, the $SO(7)$ symmetric and the $G_2$ critical, respectively, as discussed below.

\subsection{The $SO(7)$ symmetric fixed planes}
There exist three planes in which the $G_2$ symmetry is enlarged to the $SO(7)_A$, $SO(7)_B$, and $SO(7)_M$ symmetries, respectively.
They are related to each other by the triality relation as well as the RG flow patterns. The RG equations with $SO(7)$ symmetry have been investigated in literature~\cite{Lin:1998aa,Fidkowski2010}.
For the self-contentedness of this work, we summarize the
corresponding stable and critical quantum phases. 

For an $SO(7)$ plane, the coupling constants satisfy the following relation,
\beq
SO(7)_i\text{ symmetric fixed plane }P_i:~t_j=t_k=4g-2t_i,
\label{eqPm}
\eeq
where $\{i,j,k\}$ are permutations of $\{A,B,M\}$.
Within these $SO(7)$ fixed planes, the RG equations can be solved analytically.
The resulting fixed points (flows) characterize different quantum phases and phase transitions.

Taking the plane of $P_A$ as an example, the $SO(7)_A$ symmetry is explicit as seen from the reduced interaction Hamiltonian density,
\beq
\mathcal{H}_{A}=g\sum_{a\neq b}M_{ab}^R M_{ab}^L+y \sum_{a}V_a^R V_a^L.
\label{Hm}
\eeq
The RG equations in the $SO(7)_A$ symmetric plane $P_A$ are reduced to 
\beq
\frac{\text{d}g}{\text{d} l}=20 g^2 +y^2, ~~~\frac{\text{d} y }{\text{d} l}=24 g y 
\label{eqRG3}.
\eeq
The solutions to RG equations
(\ref{eqRG3}) can be classified in the following cases. 

Case (I): {\it  The fully gapped SO(8) Gross-Neveu (GN) phases.}~
There exists a stable $SO(8)$ GN \cite{Gross1974} fixed flow characterized by the interaction paramter set of
$y=t=2g \to +\infty$ and $w=0$, or, equivalently,
$t_A=t_B=t_M=\frac{4}{3}g \rightarrow +\infty$.
The Hamiltonian density along this line 
is reduced to
\beq
\mathcal{H}^{SO(8)}_\mathrm{GN}&=&g\sum_{a\neq b }M_{ab}^R M_{ab}^L+2g\sum_{a}V_a^R V_a^L=g\sum_{a\neq b}N_{ab}^R N_{ab}^L,
\label{eqso8}
\eeq
which describes the fully gapped $SO(8)$ GN phase. 
This fixed flow is at the intersection of three $SO(7)_i$
fixed planes with $i\in \{A, B, M\}$ and is invariant under the permutation of $t_i$. 
Therefore, it exhibits $SO(8)$ self-triality.

Another $SO(8)$ phase is also stable characterized by the fixed point of $y=-t=-2g \to -\infty$ and $w=0$, or, equivalently, 
$t_B=t_M=-\frac{1}{2}t_A=-\frac{4}{3}g\to -\infty$.
Along this line, the interaction Hamiltonian density is reduced to 
\beq
\mathcal{H}^{SO(8)_A}_\mathrm{GN} =g\sum_{a\neq b}M_{ab}^R M_{ab}^L-2g\sum_{a}V_a^R V_a^L.
\label{eqso8a}
\eeq
It is related to the previous $SO(8)$ phase of (\ref{eqso8}) by the chiral transformation~\cite{Shankar1981,Fidkowski2010} which keeps the current of left-movers but change that of right-movers to $\{M_{ab}^R, -V_a^R\}$.
Such a phase is also fully gapped denoted as the $SO(8)_A$ GN phase.

Case (II): {\it  The critcal SO(7) GN phase.}~
This phase is controlled by the unstable $SO(7)$ GN fixed flow characterized by 
$t=2g \to +\infty$ and $y=w=0$,
or, equivalently, 
$t_B=t_M=0$, $t_A=2g\rightarrow +\infty$.
Along this flow, the interaction Hamiltonian density is reduced to
\beq
\mathcal{H}^{SO(7)_A}_\mathrm{GN}=g\sum_{a\neq b}M_{ab}^R M_{ab}^L.\label{eqso7a}
\eeq
In such a state, 7 Majorana modes are gapped out by the GN-type interaction while 1 Majorana mode remains gapless~\cite{WU2005a,Fidkowski2010}. 
It describes a critical phase with the $SO(7)_A$ symmetry, sandwiched by two gapped $SO(8)$ phases discussed previously.  
This critical phase, denoted as the $SO(7)_A$ GN phase, has the central charge of $c=\frac{1}{2}$ attributed to the gapless Majorana mode, yielding a phase transition in the Ising universality class.
The two gapped $SO(8)$ phases are Ising dual to each other. 

Case (III): {\it  The critical $SO(8)_1$ WZW phase.}
This phase is characterized by the unstable fixed 
point of $g=t=y=w=0$, or, equivalently,
$g=t_A=t_B=t_M=0$.
Its central charge $c=4$ corresponding to the free theory of 8 Majoranas, i.e., the $SO(8)_1$ WZW model. 
It controls the gapless Luttinger liquid phase. 
Apparently, this fixed point also exhibits self-triality.

Due to the triality, RG flows and the stable phases
of the $SO(7)_B$ and $SO(7)_M$ symmetric fixed planes should exhibit the same structure as the case of $SO(7)_A$. 
Hence, three $SO(8)_i$ ($i=A,B,M$) gapped phases are related to each other by the triality. 
The phase transitions between the gapped $SO(8)$ GN phase and $SO(8)_i$ GN phases are in the Ising universality class with central charge $c=\frac{1}{2}$, controlled by $SO(7)_i$ GN fixed flows. 
The three $SO(7)_i$ GN points, as well as the three critical Ising phases are also related to each other by the triality relation. 

\begin{figure}[htbp]
\centering
\includegraphics[width=0.55\linewidth]{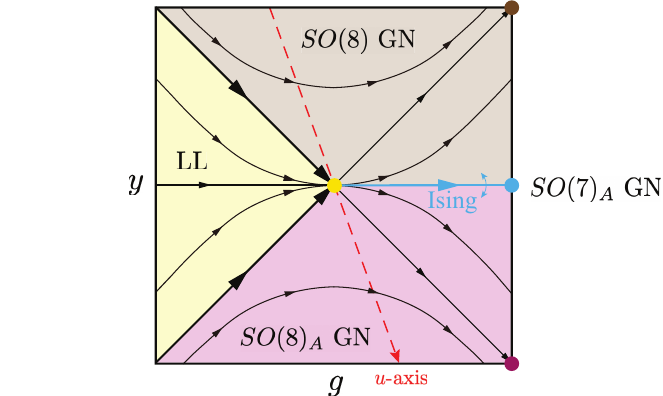}
\caption{
The RG flows and phase diagram on the $SO(7)_A$ symmetric fixed plane. 
The horizontal and vertical axes are labeled by $g$ and $y$, respectively.
Phase boundaries are marked by thick lines. 
The $SO(8)$ GN, $SO(8)_A$ GN, and $SO(7)_A$ GN fixed points (defined as $g\rightarrow +\infty$ points of the fixed flows) are marked by brown, blue, and crimson dots, respectively. 
The yellow dot stands for the free Majorana fixed point. 
``LL" denotes the Luttinger liquid phase. 
The transition between the $SO(8)$ GN phase and the $SO(8)_A$ GN phase is described by the critical $SO(7)_A$ GN phase in the Ising universality class. 
Phase structures in the $SO(7)_{B/M}$ fixed planes can be obtained via the triality mapping. 
The embedding of the $u$-axis in the lattice model phase diagram is colored in red. 
}\label{fig:planeso7}
\end{figure}

The schematic RG flows and the phase diagram on the $SO(7)_A$ symmetric fixed plane $P_A$ are shown in figure~\ref{fig:planeso7}. 
By substituting the subscript $A$ by $B/M$, the similar physics appears on $SO(7)_{B/M}$ symmetric fixed planes.
The lattice model (\ref{eq2}) with $v=0$ possesses the $SO(7)_A$ symmetry. 
As varying $u$, its phase evolution is also marked in  figure~\ref{fig:planeso7}.
In appendix~\ref{app:C}, we provide a standard bosonization analysis~\cite{Fradkin,Shankar1981,Lin:1998aa,Fidkowski2010}, which gives rise to the same phase diagram.

\subsection{The $G_2$ critical fixed planes}
Aside from the GN-type fixed flows discussed above, the RG equations (\ref{eqRG1}) have an unstable flow away from all the three $SO(7)$ symmetric fixed planes. This $G_2$ symmetric fixed flow is characterized by $g\rightarrow +\infty$ and
$t=y=w=0$, or, equivalently,
$g\rightarrow +\infty$, $t_A=t_B=t_M=0$, which exhibits the self-triality.
The corresponding interaction Hamiltonian density reads,
\beq
\mathcal{H}^{G_2}_\mathrm{TCI}=g\sum_{a\neq b}G_{ab}^R G_{ab}^L.\label{eq:tci}
\eeq
Interestingly, $g\to +\infty$ does not
completely gap out the spectra, but results in a tricritical Ising (TCI) phase~\cite{vafa1995,Hu2018} with central charge $c=\frac{7}{10}$. 
The emergent symmetry of this critical phase is not only $G_2$, but also the $\mathcal{N}=1$ spacetime supersymmetry inherited from the TCI CFT~\cite{FRIEDAN1985,QIU1986}. 

In the Majorana basis, (\ref{eq:tci}) contains only 7 of the 8 Majorana modes, and the decoupled one $\chi_0$ remains gapless. 
Hence, the total central charge is $c=\frac{7}{10}+\frac{1}{2}=\frac{6}{5}$. 
A CFT with such a central charge has a parafermion representation~\cite{Ninomia1987,Ardonne2007,Grosfeld:2009aa}, which may be realized on the edge of certain fractional quantum Hall states~\cite{Ardonne2007,Grosfeld:2009aa}.
It remains an interesting question whether some quantum Hall bilayer systems could host this model on the edge through certain modifications such as edge reconstructions~\cite{Girvin1984,MacDonald1990,Kane1994}. 
In addition, (\ref{eq:tci}) realizes the model proposed in Ref.~\cite{Hu2018} for topological superconducting edge states 
which exhibit non-Abelian Fibonacci anyonic properties.
The model of (\ref{eq3}) could be viewed as a generalization of their case.

\begin{figure}[htbp]
\centering
\includegraphics[width=0.38\linewidth]{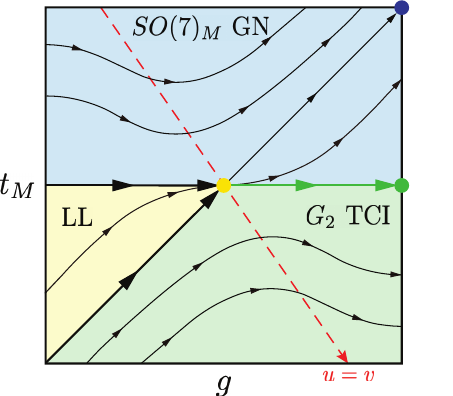}
\caption{
The RG flows and phase diagram on the $G_2$ critical fixed planes spanned by the $G_2$ TCI fixed flow and the $SO(7)_M$ critical GN fixed flow.
The horizontal and vertical axes are labeled by $g$ and $t_M$, respectively.
The $SO(7)_M$, $G_2$ TCI, and free Majorana fixed points are denoted by the indigo, green, and yellow dots, respectively. 
Phase structures in other $G_2$ critical fixed planes are similar.
The embedding of the $u=v$ line in the lattice model is colored in red.
}\label{fig:plane}
\end{figure}

The $G_2$ TCI fixed flow is special.
We can pick up any one GN-type fixed flow together with the $G_2$ TCI one to form a fixed plane. 
Consider the interaction Hamiltonian density on such a plane.
It can be rearranged as
\beq
\mathcal{H}^{G_2}_\mathrm{critical} =\tilde{g}\sum_{a\neq b}G_{ab}^R G_{ab}^L+\tilde{n}\sum_{a\neq b}\tilde{N}_{ab}^R \tilde{N}_{ab}^L,
\eeq
where $\tilde{N}_{ab}$ are the generators of the emergent symmetry on the GN fixed flow. For example, for the $SO(8)_A$ GN fixed flow, $\{\tilde{N}_{ab}^R\}=\{M_{ab}^R,-V_{c}^R\}$ and $\{\tilde{N}_{ab}^L\}=\{M_{ab}^L,V_{c}^L\}$ are generators of the $SO(8)_A$ symmetry. 
Since the emergent symmetry is enlarged including the $G_2$ as a subgroup, the closure of Lie algebra ensures the RG flow to remain in-plane. 
The corresponding fixed points (flows) and phases are the GN fixed flow controlling the gapped or critical phase with emergent symmetry, the $G_2$ TCI fixed flow controlling the $G_2$ TCI phase, and the free Majorana fixed point contronlling the Luttinger liquid phase. 
As an example, we choose the GN fixed flow as the $SO(7)_M$ critical one. Correspondingly the fixed plane has $t_A=t_B=0$, on which the RG equations are reduced to 
\beq
\frac{\mathrm{d}g}{\mathrm{d}l} =16 g^2 +t_M^2, ~~~\frac{\mathrm{d}t_M}{\mathrm{d}l} =16 g t_M +2 t_M^2.
\label{eq:RGso7_M}
\eeq
It indeed possesses the $SO(7)_M$
critical fixed flow of $t_A=t_B=0$ and $t_M=2g \to +\infty$, or, equivalently, $t=\frac{1}{3}y=-\frac{1}{\sqrt{6}}w=\frac{1}{2}g\to +\infty$. 
Along the $SO(7)_M$ critical flow, the interaction Hamiltonian density is reduced to
\beq
\mathcal{H}_\text{GN}^{SO(7)_M}&=&g\sum_{a\neq b}G_{ab}^R G_{ab}^L+\frac{g}{2}\sum_{a}T_a^R T_a^L+
\frac{3 g}{2} \sum_{a}V_a^R V_a^L-\frac{\sqrt 3 g}{2} \sum_{a}
\left(T_a^R V_a^L+V_a^R T_a^L\right)\nn\\
&=&g\sum_{a\neq b}M_{ab}^{\prime\prime R} M_{ab}^{\prime\prime L}
.
\eeq
The schematic RG flows and the phase diagram on this $G_2$ critical fixed planes are shown in figure~\ref{fig:plane}.
A complete list of fixed flows is given in table~\ref{tab:1}.

\begin{table}[htbp]
\centering
\begin{tabular}{ccc|c|c}
\hline
$t_A/g$ & $t_B/g$ & $t_M/g$ & Symmetry & Fixed plane\\
\hline
$4/3$ & $4/3$ & $4/3$ & $SO(8)$ & $P_A,~P_B,~P_M$\\
\hline
$8/3$ & $-4/3$ & $-4/3$ & $SO(8)_A$ & $P_A$\\
$2$ & $0$ & $0$ & $SO(7)_A$ & $P_A$\\
\hline
$-4/3$ & $8/3$ & $-4/3$ & $SO(8)_B$ & $P_B$\\
$0$ & $2$ & $0$ & $SO(7)_B$ & $P_B$\\
\hline
$-4/3$ & $-4/3$ & $8/3$ & $SO(8)_M$ & $P_M$\\
$0$ & $0$ & $2$ & $SO(7)_M$ & $P_M$\\
\hline
$0$ & $0$ & $0$ & $G_2$ SUSY & Any\\
\hline
\end{tabular}
\caption{Fixed flows ($g\neq 0$), the associated symmetries, and the corresponding fixed planes. $P_i$ ($i=A,B,M$) represents the $SO(7)_i$ symmetric fixed plane. 
The $G_2$ TCI fixed flow has emergent supersymmetry endowed by the TCI CFT. 
In the last row, ``Any" means a plane spanned by the $G_2$ TCI fixed flow and anyone of the GN fixed flows is a fixed plane.
}\label{tab:1}
\end{table}

\subsection{Phase diagrams in fixed bodies}

The transitions between the $SO(8)$ GN phase and each of the $SO(8)_i$ GN phase is revealed by the Ising duality in the corresponding $SO(7)$ fixed plane. 
Nevertheless, phase transitions between any two of these three $SO(8)_i$ GN phases are more involved, since different stable GN fixed flows are not in the same fixed plane. 
To investigate this problem, we find the following relation based on the RG equations (\ref{eqRG2}),
\beq
\frac{\mathrm{d}\tilde{t}}{\mathrm{d}l}=16 g\tilde{t},
\eeq
with $\tilde{t}=t_A+t_B+t_M$.
Hence, during the evolution $\tilde{t}=0$ is maintained.
The fixed body spanned by $g$ and $\tilde{t}=0$ includes all the three $SO(8)_i$ GN fixed flows exhibiting the triality. 

\begin{figure}[htbp]
\centering
\includegraphics[width=0.8\linewidth]{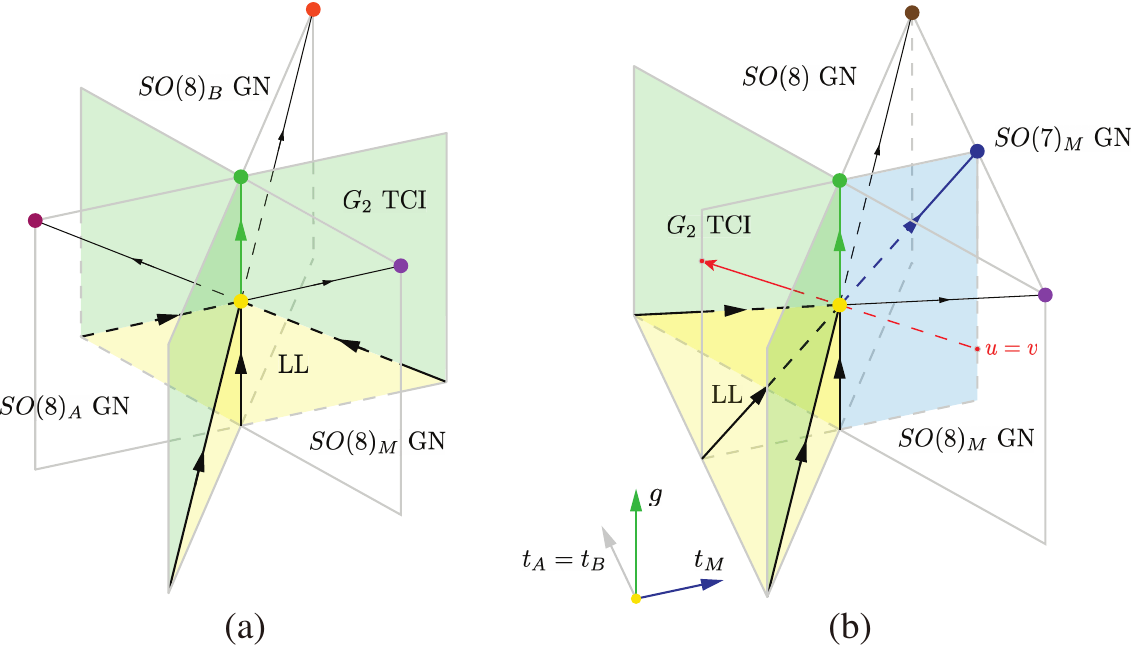}
\caption{The phase diagrams in the fixed bodies (a) $\tilde{t}=t_A+t_B+t_M=0$ and (b) $t_A=t_B$. 
The fixed flows are marked as arrows, and the critical surfaces are colored. 
In (a), $SO(8)_A$ GN, $SO(8)_B$ GN, $SO(8)_M$ GN, $G_2$ TCI, and free Majorana fixed points are denoted by crimson, orange, purple, green, and yellow dots, respectively. 
The transition between each two of the three gapped $SO(8)_i$ GN phases is through either the $G_2$ TCI phase or the Luttinger liquid (LL) critical phase. The triality relation is manifest in the $D_3$ symmetry of the phase diagram. 
In (b), the $SO(8)$ GN, $SO(7)_M$ GN, $SO(8)_M$ GN, $G_2$ TCI, and free Majorana fixed points are denoted by brown, indigo, purple, green, and yellow dots, respectively. 
The transition between the $SO(8)$ GN and the $SO(8)_M$ GN phases is through the critical $SO(7)_M$ GN phase in the Ising universality class. 
The embedding of the $u=v$ axis in the lattice model phase diagram is shown in red, whose positive and negative semi-axes are embedded in the $G_2$ TCI and the critical $SO(7)_M$ GN phases, respectively.}
\label{fig:body}
\end{figure}

The phase diagram in fixed body $\tilde{t}=0$ is shown in figure~\ref{fig:body} (a), which exhibits the $D_3$ symmetry as a consequence of the triality among the three $SO(8)_i$ phases. 
The $G_2$ TCI fixed flow and the free Majorana fixed point also lie in this fixed body, as indicated in table~\ref{tab:1}. 
The phase diagram shows that the transition between any two of the three $SO(8)_i$ GN phases is described by either the Luttinger liquid phase with central charge $c=4$, or, the $G_2$ TCI phase with central charge $c=\frac{6}{5}$, both of which are actually multi-critical. 
This coincides with the phenomenon of multiversality that the phase transitions between two phases can have more than one universality classes~\cite{Bi2019,Abhishodh:2023}. 

The RG equations  (\ref{eqRG2}) also show that if any two of $t_A=t_B=t_M$ are equal, this relation is also maintained during the RG evolution. 
Hence, there exist another class of three fixed bodies given by $t_A=t_B$, $t_B=t_M$, and $t_M=t_A$, respectively, which are related by triality. 
As an example, the phase diagram in the fixed body $t_A=t_B$ is shown in figure~\ref{fig:body} (b), in which the $SO(7)_M$ fixed plane $P_M$ spanned by three GN fixed flows is embedded. 

\section{Embedding of the lattice model}
\label{sect:embedding}

Neither triality nor the spacetime supersymmetry survives in the 1D lattice Hubbard-type model of (\ref{eq2}). 
It is natural to ask the reason behind such a loss. 
This is equivalent to investigate the embedding of the lattice model phase diagram into that of the continuum model.
The phase diagram of the lattice Hamiltonian (\ref{eq2}) is spanned by $u$- and $v$-axis, along which the $SO(7)_A$ and $SO(7)_B$ symmetries are restored, respectively. 
Therefore, the $u/v$-axis must be a straight line passing the origin point in the $SO(7)_{A/B}$ fixed plane $P_{A/B}$ in the continuum model phase diagram. 

A possible embedding of the $u$-axis is shown in figure~\ref{fig:planeso7}.
According to (\ref{eqPm}),
in the parameter space of the continuum model (\ref{eq4}),
the $u$-axis of the lattice model 
(\ref{eq2}) possessing the
$SO(7)_A$ symmetry is parameterized
as
\beq
u\text{-axis}:~t_A=\left(2-\frac{k}{2}\right)g, ~t_B=kg, ~t_M=kg,
\eeq
where $k$ is a constant to be determined. 
Based on the triality mapping, the $v$-axis  possessing the
$SO(7)_B$ symmetry is obtained by switching 
$A$ and $B$ and fixing $M$, hence, 
it is parameterized as
\beq
v\text{-axis}:~t_A=kg,~t_B=\left(2-\frac{k}{2}\right)g,~t_M=kg
\eeq
Therefore, the phase diagram of the lattice model (\ref{eq2}) spanned by $u$-axis and $v$-axis is given by
\beq
P_\mathrm{Lat}:~t_A+t_B=\left(2+\frac{k}{2}\right)g,~t_M=kg.
\label{eq:plat}
\eeq

To determine the value of $k$, the $u=v$ axis 
is checked in the lattice model phase diagram.
The lattice Hubbard type interaction (\ref{eq2}) is invariant under the transformation
$\chi_0(i) H_\mathrm{int} \chi_0(i) =H_\mathrm{int}$~\cite{Gao_2020}.
Hence, $\chi_0$ cannot appear in the local interaction of $H_\mathrm{int}$.
In other words, $\chi_0$ decouples from other Majorana modes, and hence, the system should be gapless along the entire $u=v$ axis. 
In $P_\mathrm{Lat}$, the $u=v$ axis reads
\beq
u=v:~t_A=t_B=\left(1+\frac{k}{4}\right)g,~t_M=kg,
\eeq
which lies in the fixed body $t_A=t_B$. 
To have both its positive and negative semi-axes embedded in gapless phases, the only choice of $k$ is $k=-4$. 

The reasoning why the $SO(7)_M$ is absent in the lattice Hamiltonian (\ref{eq2}) is presented below. 
The equation of $SO(7)_M$ fixed plane $P_M$
for the continuum Hamiltonian density
is given by $t_A=t_B=4g-2t_M$. 
However, the $u$-$v$ plane (\ref{eq:plat}) of $P_\mathrm{Lat}$ with $k=-4$ substituted, i.e., $t_A+t_B=0$ and $t_M=-4g$
is incompatible with the $SO(7)_M$ 
fixed plane $P_M$, except the trivial case of $g=t_A=t_B=t_M=0$, i.e. $u=v=0$. 
The triality among three $SO(7)$ symmetries in the continuum model is broken to the duality of $SO(7)_A$ and $SO(7)_B$ on the lattice. 
This is a reminiscence of triality upon lattice regularization. 

We need to figure out how the lattice Hamiltonian (\ref{eq2}) with $u=v>0$ and $<0$ matches the $G_2$ critical TCI fixed point and $SO(7)_M$ critical ones, respectively. 
For the $G_2$ TCI fixed flow (\ref{eq:tci}),
the case of $u=v>0$ of the lattice Hamiltonian should match the case $g>0$ of the continuum one. 
Thus the positive semi-axis of $u=v$ should be embedded in the $G_2$ critical phase, and then naturally the negative semi-axis is embedded 
in the $SO(7)_M$ critical phase, as shown in figures~\ref{fig:plane} and \ref{fig:body} (b).
Consequently, in the lattice model interaction with $u=v>0$ flows to the $G_2$ critical TCI fixed point, and that of $u=v<0$ flows to the $SO(7)_M$ critical fixed point.

\begin{figure}[htbp]
\centering
\includegraphics[width=1\linewidth]{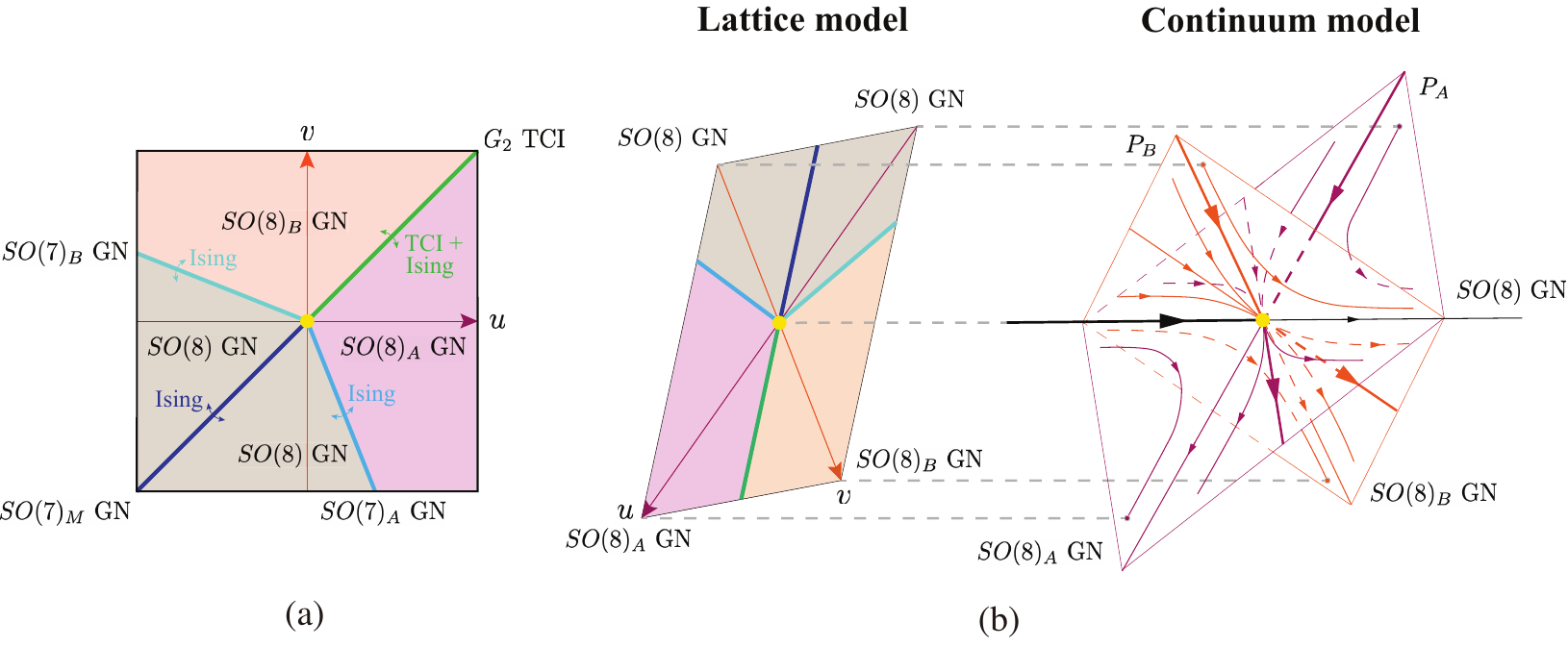}
\caption{(a) The lattice model phase diagram. There are totally three gapped phases, belonging to $SO(8)_A$ GN, $SO(8)_B$ GN, and $SO(8)$ GN phases in the continuum model. The phase transition between $SO(8)_A$ GN and $SO(8)_B$ GN phases, marked as the green segment, is located at $u=v>0$, and described by $G_2$ TCI phase in the continuum model. The phase transition between $SO(8)_A$ ($SO(8)_B$) GN and $SO(8)$ GN phases, represented by the blue (turquoise) segment, is located at $u=-7v$ ($v=-7u$) and belongs to the critical $SO(7)_A$ ($SO(7)_B$) GN phase in the continuum model. Inside the $SO(8)$ GN phase there is a branch cut colored in indigo, on which the system becomes gapless. It is located at $u=v<0$ and described by the critical $SO(7)_M$ GN phase in the continuum model. 
(b) Embedding of the lattice model phase diagram into the continuum model phase diagram. RG flows of the continuum model in body $4g-2t_A-2t_B+t_M=0$ (not a fixed body) spanned by $SO(7)_{A}$ and $SO(7)_B$ symmetric fixed planes ($P_{A}$ and $P_B$) are shown in the right. On each $SO(7)$ symmetric fixed plane the RG flows are the same as figure~\ref{fig:plane} (a) ensured by the triality. The phase boundaries are marked as thick segments. $P_A$ and $P_B$, as well as RG flows on them, are colored in crimson and orange, respectively. 
The lattice model phase diagram is shown in the left part. The $u/v$-axis is embedded in $P_{A/B}$ as a straight line.}
\label{fig:lattice}
\end{figure}

The phase diagram of the lattice model
Hamiltonian (\ref{eq2}) is shown in figure~\ref{fig:lattice} (a), supplemented by a numerical integration of the RG equations on the embedded lattice model phase diagram $P_\mathrm{Lat}:~t_A+t_B=0,~t_M=-4g$ (see appendix~\ref{app:D} for details). The Ising transition line between $SO(8)_A$ ($SO(8)_B$) GN phase and $SO(8)$ GN phase is located at $u=-7v$ ($v=-7u$), which perfectly agrees with the mean-field analysis of the lattice model~\cite{Gao_2020}. 
The embedding of the lattice phase diagram is shown in figure~\ref{fig:lattice} (b).
The $u$-axis possesses the $SO(7)_A$ symmetry.
Compared to the phase diagram of the continuum model shown in figure~\ref{fig:planeso7}, the sign of $g$ in the continuum Hamiltonian density follows that of $u$, while that of $y$ is opposite to $u$. 
Hence, the positive $u$-axis is embedded in the $SO(8)_A$ GN phase, and its negative axis is embedded in the $SO(8)$ GN phase.
By the triality mapping, the positive $v$-axis is embedded in the $SO(8)_B$ GN phase, and the negative $v$-axis is also in the $SO(8)$ GN phase.

For the $u=v$ axis, its negative semi-axis ($u=v<0$) embedded in the critical $SO(7)_M$ GN phase is a branch cut lying inside the gapped $SO(8)$ GN phase. 
Its gaplessness is due to the decoupled free Majorana $\chi_0$, which is consistent with the analysis of the $u=v$ axis in the lattice model~\cite{Gao_2020}. 
This criticality sandwiched in the same phase is an example of the unnecessary criticality~\cite{Bi2019,Abhishodh:2023}.
On the other hand, the positive $u=v$ semi-axis, embedded in the $G_2$ critical phase, describes the transition between $SO(8)_A$ GN and $SO(8)_B$ GN phases in the lattice model. 
Such a transition on lattice inherits the universality class and the central charge $c=\frac{7}{10}+\frac{1}{2}$ of the $G_2$ TCI phase in the continuum. 

The spacetime supersymmetry in the $G_2$ TCI phase also has a reminiscence on the lattice. For a CFT with $\mathcal{N}=1$ spacetime supersymmetry, the primary fields come in pairs. Within each pair the two primary fields have scaling dimension differed by $\frac{1}{2}$, and hence can be viewed as the supersymmetry partner of each other. For example, for the TCI CFT described by (\ref{eq:tci}), the stress-energy tensor for right-mover $T_B^R\sim C_{abcd}\chi_{a+1}^R\chi_{b+1}^R\chi_{c+1}^R\chi_{d+1}^R$ with scaling dimension $2$ has a supersymmetry partner $T_F^R\sim C_{abc}\chi_{a+1}^R\chi_{b+1}^R\chi_{c+1}^R$ with scaling dimension $\frac{3}{2}$~\cite{vafa1995} (this operator is also called $\epsilon^{\prime\prime}$ in CFT literature). Typically, a lattice model with low energy effective theory governed by a supersymmetric CFT exhibits quantum mechanical supersymmetry~\cite{Fendley2003,LiSUSY}. Similar to the supersymmetric CFT, in a lattice model with $\mathcal{N}=1$ quantum mechanical supersymmetry, the eigenstates of the Hamiltonian $H_\mathrm{SUSY}$ also come in pairs, within which the two states have different fermion parity and related to each other by a supercharge operator $Q$ satisfying $H_\mathrm{SUSY}=Q^2$. This phenomenon particularly resembles the $G_2$ invariant lattice model (\ref{eq2}) when $u=v$, which exhibits the degeneracy between bosonic and fermionic states on each site~\cite{Gao_2020}.
The ``supercharge operator" switching bosonic and fermionic states on site $i$ is just the Majorana operator $\chi_0(i)$. 
However, this is not an authentic supersymmetry since $H^{G_2}_\text{{int}}\neq(\sum_i\chi_0(i))^2$, and hence does not satisfy the definition of the quantum mechanical supersymmetry. 
Nevertheless, this degeneracy between on-site bosonic and fermionic states can still be viewed as the lattice reminiscence of the spacetime supersymmetry in the continuum model.

Before closure, we briefly comment the fate of spacetime supersymmetry at $u=v$ in higher dimensions. In spatial dimension $d>1$, sufficiently strong $u$ and $v$ will drive $G_2$ symmetry to be spontaneously broken to $SU(3)$ or $SU(2)\times U(1)$~\cite{Gao_2020}, with order parameters transformed in the 7-dimensional vector representation or both 7-dimensional vector and 14-dimensional adjoint representations, respectively. Upon Hubbard-Stratonovich transformation, scalar boson fields are introduced and coupled to the Majorana fermions through a Yukawa-type interaction. We first consider the case of $u=v<0$, where totally 21 scalar boson fields are introduced in the Hubbard-Stratonovich transformation to capture the spontaneous breaking of $G_2$ to $SU(2)\times U(1)$. 7 of the bosons are transformed in the 7-vector representation of $G_2$, while the other 14 are in the adjoint representation. The adjoint bosons do not have fermionic partners, and hence spacetime supersymmetry are not expected.

For $u=v>0$, 7 bosons are introduced transformed as the 7-vector of $G_2$. When $u<u_c$, these bosons are gapped, corresponding to the $G_2$ symmetric phase where all the 8 Majoranas are gapless. When $u>u_c$, condensation of the bosons spontaneously breaks $G_2$ down to $SU(3)$. Consequently, 6 out of 8 Majoranas are gapped, composing a 3-component complex fermion transformed in the 3-spinor representation of $SU(3)$~\cite{Gao_2020}. The rest 2 gapless Majoranas always include $\chi_0$ which remains decoupled and gapless on the entire $u=v$ line. At the critical point $u=u_c$, all bosons and Majoranas become gapless simultaneously, while 7 Majoranas $\chi_m$, $1\le m\le 7$, are transformed in the same 7-vector representation of $G_2$ as the 7 real bosons, signaling a plausible supersymmetry structure. A necessary condition for an authentic spacetime supersymmetry is that the low energy Majoranas must be relativistic in order to be supersymmetry partners of the scalar bosons, which can be realized in topological superconductors~\cite{Li2012,Grover2014,Pan2024}. In this case, upon Hubbard-Stratonovich transformation, the critical theory for $u=u_c$ is a Gross-Neveu-Yukawa theory~\cite{Gross1974} with 7-flavor scalar bosons and Majorana fermions (decoupled $\chi_0$ omitted):
\beq
\mathcal{L}=\bar{\chi}_m\big(i\gamma^\mu\partial_\mu\delta_{mn}-i\tilde{C}_{mnp}\varphi_p\big)\chi_{n}+U(\varphi).
\eeq
Here $\gamma^\mu$ are gamma matrices, $\tilde{C}_{mnp}=C_{(m-1)(n-1)(p-1)}$, and $U(\varphi)$ is the quartic self-interaction of the scalar boson $\varphi_m$. Details of the Hubbard-Stratonovich transformation is given in appendix~\ref{app:A}. Spacetime supersymmetry requires the existence of superpotential $W$ satisfying~\cite{Bilal2001}
\beq
\frac{1}{2}\frac{\partial^2 W}{\partial\varphi_m\partial\varphi_n}=i\tilde{C}_{mnp}\varphi_p,~~~ \left(\frac{\partial W}{\partial\varphi_m}\right)^2=U(\varphi).
\eeq
However, the former equality is self-conflicting since $\tilde{C}_{mnp}$ is fully anti-symmetric. Therefore, we conclude that supersymmetry cannot emerge at $u=v$ of the lattice model (\ref{eq2}) in higher dimensions.

\section{Conclusions}
\label{sect:conclusions}

In summary, we investigate the low energy effective theory of the minimal $G_2$ symmetric lattice model~\cite{Gao_2020} in (1+1)D by an one-loop RG method. 
The $SO(8)$ triality, which is absent in the lattice model due to the on-site Hubbard-type interactions, emerges naturally in the low energy effective theory as a triality mapping among three different $SO(7)$ symmetries, and consequently three gapped phases as well as three Ising critical phases. 
The phase transitions between phases related by the triality relation are unified by either a Luttinger 
liquid phase, or a $G_2$ symmetric multi-critical phase. This $G_2$ multi-critical phase is described by a direct product of Ising and TCI CFTs with central charge $c=\frac{1}{2}+\frac{7}{10}=\frac{6}{5}$, and possesses emergent spacetime supersymmetry carried by the TCI CFT. 
When the continuum model is regularized on a lattice, the triality among three $SO(7)$ symmetries is broken to the duality of two $SO(7)$ symmetries~\cite{Gao_2020}, eliminating the third one. 
The spacetime supersymmetry is also degraded to a degeneracy pattern between the on-site bosonic and fermionic states, instead of an authentic quantum mechanical supersymmetry. However, these reminiscences in ultraviolet still imply clues of exotic structures in infrared, as explored in the current work.

\acknowledgments
ZQG acknowledges D.-H. Lee, H. Yang, Y.-Q. Wang, Z. Han, and J.-C. Feng for helpful discussions.
CW is supported by the National Natural Science Foundation of China under the Grant No. 12234016 and No. 12174317.
This work has been supported by the New Cornerstone Science Foundation. 

\bibliography{g2rg}
\bibliographystyle{JHEP}

\clearpage

\appendix
\section{Fermion bilinears and majorana operators in the lattice model}\label{app:A}

Define four by four gamma matrices as
\beq
\Gamma^1 =\begin{pmatrix}
0 & -iI\\
iI & 0
\end{pmatrix},\quad
\Gamma^{2,3,4} =\begin{pmatrix}
\vec{\sigma} & 0\\
0 & -\vec{\sigma}
\end{pmatrix},\quad
\Gamma^5 =\begin{pmatrix}
0 & I\\
I & 0
\end{pmatrix},
\eeq
where $I$ and $\vec{\sigma}$ are two by two identity and Pauli matrices. These five gamma matrices form an $Sp(4)$ vector. The generators of $Sp(4)$ are defined as $\Gamma^{ab}=-\frac{i}{2}[\Gamma^a, \Gamma^b]$, and the charge conjugation operator is defined as $R=\Gamma^1\Gamma^3$. Therefore, the femion bilinears on each site $i$, $N_{ab}(i)$ with $0\le a, b\le 7$, spanning an $SO(8)$ algebra reads
\beq
&&N_{ab}(i)=-\frac{1}{2}\psi^\dagger(i) \Gamma^{ab} \psi(i),\quad 1\le a, b\le 5,\\
&&N_{a7}(i)=\frac{1}{2}\psi^\dagger(i) \Gamma^a \psi(i),\quad 1\le a\le 5,\\
&&N_{06}(i)=\frac{1}{2}\psi^\dagger(i) \psi(i)-1,\\
&&N_{0a}(i)+iN_{a6}(i)=-\frac{i}{2}\text{sgn}(i)\psi^\dagger (i) \Gamma^a R \psi^\dagger(i),\quad 1\le a\le 5,\\
&&N_{07}(i)-iN_{67}(i)=\frac{1}{2}\text{sgn}(i)\psi^\dagger (i) R \psi^\dagger(i),
\eeq
where sgn$(i)=\pm 1$ is staggered on two sublattices of the bipartite lattice. The Majorana operators, transformed under the vector representation of $SO(8)$, are defined as
\bea
&\chi_1(i)=e^{-i\text{sgn}(i)\pi/4} \psi_\frac{1}{2}(i)+e^{i\text{sgn}(i)\pi/4}\psi^\dagger_\frac{1}{2}(i),\\
&\chi_2(i)=-e^{i\text{sgn}(i)\pi/4} \psi_\frac{1}{2}(i)-e^{-i\text{sgn}(i)\pi/4}\psi^\dagger_\frac{1}{2}(i),\\
&\chi_3(i)=e^{-i\text{sgn}(i)\pi/4} \psi_{-\frac{1}{2}}(i)+e^{i\text{sgn}(i)\pi/4}\psi^\dagger_{-\frac{1}{2}}(i),\\
&\chi_4(i)=-e^{i\text{sgn}(i)\pi/4} \psi_\frac{3}{2}(i)-e^{-i\text{sgn}(i)\pi/4}\psi^\dagger_\frac{3}{2}(i),\\
&\chi_5(i)=e^{i\text{sgn}(i)\pi/4} \psi_{-\frac{1}{2}}(i)+e^{-i\text{sgn}(i)\pi/4}\psi^\dagger_{-\frac{1}{2}}(i),\\
&\chi_6(i)=e^{-i\text{sgn}(i)\pi/4} \psi_{-\frac{3}{2}}(i)+e^{i\text{sgn}(i)\pi/4}\psi^\dagger_{-\frac{3}{2}}(i),\\
&\chi_7(i)=e^{i\text{sgn}(i)\pi/4} \psi_{-\frac{3}{2}}(i)+e^{-i\text{sgn}(i)\pi/4}\psi^\dagger_{-\frac{3}{2}}(i)\\
&\chi_0(i)=e^{-i\text{sgn}(i)\pi/4} \psi_\frac{3}{2}(i)+e^{i\text{sgn}(i)\pi/4}\psi^\dagger_\frac{3}{2}(i),\label{eq:31}
\eea
where $\chi_{1,2,...,7}(i)$ form a $G_2$ vector while $\chi_0(i)$ is a $G_2$ singlet. In fact, $\chi_{1,2,...,7}(i)$ also form a vector of $SO(7)_M$ spanned by $i\chi_m(i)\chi_n(i)$ with $1\le m,n\le 7$, which is absent in the lattice model. The $G_2$ weights of these Majorana fermions are given in table~\ref{tab:weight}. As a comparison, the weight of bosonic and fermionic states are also listed.

\begin{table}[!htbp]
\begin{tabular}{c|c|c|c}
\hline
Weight $\mu$ & Bosonic vector & Fermionic vector & Majorana vector\\
\hline
$(\frac{\sqrt{3}}{2},\frac{1}{2})$ & $\psi_{\frac{1}{2}}^\dag\psi_{-\frac{3}{2}}^\dag\left| 0\right>$ & $\psi_{\frac{3}{2}}^\dag\psi_{\frac{1}{2}}^\dag\psi_{-\frac{3}{2}}^\dag\left| 0\right>$ & $\frac{1}{2}(\chi_1+i\chi_2)$\\
\hline
$(0,1)$ & $\psi_{\frac{3}{2}}^\dag\psi_{\frac{1}{2}}^\dag\psi_{-\frac{1}{2}}^\dag\psi_{-\frac{3}{2}}^\dag\left| 0\right>$ & $\psi_{\frac{1}{2}}^\dag\psi_{-\frac{1}{2}}^\dag\psi_{-\frac{3}{2}}^\dag\left| 0\right>$ & $\frac{1}{2}(\chi_6-i\chi_7)$\\
\hline
$(-\frac{\sqrt{3}}{2},\frac{1}{2})$ & $\psi_{-\frac{1}{2}}^\dag\psi_{-\frac{3}{2}}^\dag\left| 0\right>$ & $\psi_{\frac{3}{2}}^\dag\psi_{-\frac{1}{2}}^\dag\psi_{-\frac{3}{2}}^\dag\left| 0\right>$ & $\frac{1}{2}(\chi_3-i\chi_5)$\\
\hline
$(-\frac{\sqrt{3}}{2},-\frac{1}{2})$ & $\psi_{\frac{3}{2}}^\dag\psi_{-\frac{1}{2}}^\dag\left| 0\right>$ & $\psi_{-\frac{1}{2}}^\dag\left| 0\right>$ & $\frac{1}{2}(\chi_1-i\chi_2)$\\
\hline
$(0,-1)$ & $\left| 0\right>$ & $\psi_{\frac{3}{2}}^\dag\left| 0\right>$ & $\frac{1}{2}(\chi_6+i\chi_7)$\\
\hline
$(\frac{\sqrt{3}}{2},-\frac{1}{2})$ & $\psi_{\frac{3}{2}}^\dag\psi_{\frac{1}{2}}^\dag\left| 0\right>$ & $\psi_{\frac{1}{2}}^\dag\left| 0\right>$ & $\frac{1}{2}(\chi_3+i\chi_5)$\\
\hline
$(0,0)$ & $\frac{1}{\sqrt{2}}(\psi_{\frac{3}{2}}^\dag\psi_{-\frac{3}{2}}^\dag+\psi_{\frac{1}{2}}^\dag\psi_{-\frac{1}{2}}^\dag)\left| 0\right>$ & $\frac{1}{\sqrt{2}}(\psi_{-\frac{3}{2}}^\dag\pm i\psi_{\frac{3}{2}}^\dag\psi_{\frac{1}{2}}^\dag\psi_{-\frac{1}{2}}^\dag)\left| 0\right>$ & $\chi_4$\\
\hline
\hline
Weight $\mu$ & Bosonic singlet & Fermionic singlet & Majorana singlet\\
\hline
(0,0) & $\frac{1}{\sqrt{2}}(\psi_{\frac{3}{2}}^\dag\psi_{-\frac{3}{2}}^\dag-\psi_{\frac{1}{2}}^\dag\psi_{-\frac{1}{2}}^\dag)\left| 0\right>$ & $\frac{1}{\sqrt{2}}(\psi_{-\frac{3}{2}}^\dag\mp i\psi_{\frac{3}{2}}^\dag\psi_{\frac{1}{2}}^\dag\psi_{-\frac{1}{2}}^\dag)\left| 0\right>$ & $\chi_0$
 \\
\hline
\end{tabular}
\caption{Bosonic states, fermionic states, and Majorana operators classified by $G_2$ weight. The $\pm$ signs in the $(0,0)$ weight states are staggered on sublattices. The site label is omitted for clarity.}\label{tab:weight}
\centering
\end{table}

In this Majorana basis, the lattice Hamiltonian reads (constant term neglected)
\beq
H_0&=&\frac{it_\text{hop}}{2}\Big(\sum_i\chi_0(i)\chi_0(i+1)+\chi_1(i)\chi_1(i+1)+\chi_3(i)\chi_3(i+1)+\chi_6(i)\chi_6(i+1)\nn\\
&\quad &\quad\quad\quad -\chi_2(i)\chi_2(i+1)-\chi_4(i)\chi_4(i+1)-\chi_5(i)\chi_5(i+1)-\chi_7(i)\chi_7(i+1)\Big)+h.c.\nn\\
&&\\
H_{\text{int}}^{G_2}&=&-\frac{u+v}{32}\sum_i C_{abcd}\chi_{a+1}(i)\chi_{b+1}(i)\chi_{c+1}(i)\chi_{d+1}(i)\nn\\
&\quad &+\frac{u-v}{8}\sum_i \chi_{0}(i)C_{abc}\chi_{a+1}(i)\chi_{b+1}(i)\chi_{c+1}(i),
\label{eq:34}
\eeq
where the structure constant of octonions $C_{abc}$ is chosen as $C_{031} = C_{052} = C_{064} = C_{126} = C_{154} = C_{243} = C_{365} = 1$ with dual tensor $C_{ijkl}=\frac{1}{6}\epsilon_{ijklabc}C_{abc}$. When $u=v$, the interaction Hamiltonian reads
\beq
H_{\text{int}}^{G_2}=\frac{3}{2}u\sum_i G_{ab}(i)^2=-\frac{u}{16}\sum_i C_{abcd}\chi_{a+1}(i)\chi_{b+1}(i)\chi_{c+1}(i)\chi_{d+1}(i),\label{eq:33}
\eeq
from which it can be seen that only 7 out of 8 Majoranas, except for $\chi_0(i)$, appear in this Hamiltonian. It can be rearranged as
\beq
H_{\text{int}}^{G_2}&=&-\frac{u}{16}\sum_i\big(iC_{mab}\chi_{a+1}(i)\chi_{b+1}(i)\big)\big(iC_{mcd}\chi_{c+1}(i)\chi_{d+1}(i)\big),
\eeq
where identity $C_{mab}C_{mcd}=-C_{abcd}+\delta_{ac}\delta_{bd}-\delta_{ad}\delta_{bc}$ is used. Therefore, a Hubbard-Stratonovich transformation generates a Yukawa coupling between Majorana $\chi_a$ and scalar boson $\phi_a$ as $iC_{abc}\chi_{a+1}\chi_{b+1}\phi_{c+1}$.

\section{RG equations}\label{app:B}

The current algebras read (constant terms neglected since they do not contribute to the RG equations)
\beq
G_{ab}^R(z)G_{cd}^R(w)&\sim &\frac{1}{z-w}\left(\frac{2}{3}\delta_{ac}G_{bd}^R(w)+\frac{2}{3}\delta_{bd}G_{ac}^R(w)-\frac{2}{3}\delta_{ad}G_{bc}^R(w)-\frac{2}{3}\delta_{ac}G_{bd}^R(w)\right.\label{eqGG}\\
&\quad &+\left. \frac{1}{6}C_{bcam}G_{md}^R(w)+\frac{1}{6}C_{dabm}G_{mc}^R(w)-\frac{1}{6}C_{bcdm}G_{ma}^R(w)-\frac{1}{6}C_{dacm}G_{mb}^R(w)\right)\nn\\
G_{ab}^R(z)T_c^R(w)&\sim &\frac{1}{z-w}\left(\frac{2}{3}\delta_{bc}T_a^R(w)-\frac{2}{3}\delta_{ac}T_b^R(w)+\frac{1}{3}C_{abcd}T_d^R(w)\right),\\
G_{ab}^R(z)V_c^R(w)&\sim &\frac{1}{z-w}\left(\frac{2}{3}\delta_{bc}V_a^R(w)-\frac{2}{3}\delta_{ac}V_b^R(w)+\frac{1}{3}C_{abcd}V_d^R(w)\right),\\
T_a^R(z)T_b^R(w)&\sim &\frac{1}{z-w}\left(-G_{ab}^R(w)+\frac{1}{\sqrt{3}} C_{abc}T_c^R(w)\right),\\
T_a^R(z)V_b^R(w)&\sim &\frac{1}{z-w}\left(-\frac{1}{\sqrt{3}} C_{abc}V_c^R(w)\right),\\
V_a^R(z)V_b^R(w)&\sim &\frac{1}{z-w}\left(-G_{ab}^R(w)-\frac{1}{\sqrt{3}} C_{abc}T_c^R(w)\right),
\eeq
and for left-handed currents the current algebra is the same, except for $z$ substituted by $\bar{z}$. The WZW level is irrelevant to the one-loop RG of interaction Hamiltonian density, so it is simply dropped. Consider operator product expansions (OPE) showing up in the calculation of one-loop RG as follows.

1.
\beq
G_{ab}^R(z)G_{ab}^L(\bar{z})G_{cd}^R(w)G_{cd}^L(\bar{w})\sim \frac{1}{|z-w|^2}\frac{4}{9}\left(\delta\cdot G^R+\frac{1}{4}C\cdot G^R\right)\left(\delta\cdot G^L+\frac{1}{4}C\cdot G^L\right),
\eeq
where shorthands $\delta\cdot G$ and $C\cdot G$ are used to represent four $\delta$-terms and four $C$-terms in (\ref{eqGG}) for simplicity. Consider $(\delta\cdot G^R)(\delta\cdot G^L)$ first.
\beq
(\delta\cdot G^R)(\delta\cdot G^L)&=&(\delta_{ac}G_{bd}^R+\delta_{bd}G_{ac}^R-\delta_{ad}G_{bc}^R-\delta_{ac}G_{bd}^R)(\delta_{ac}G_{bd}^L+\delta_{bd}G_{ac}^L-\delta_{ad}G_{bc}^L-\delta_{ac}G_{bd}^L)\nn\\
&=&-2\delta_{ad}\delta_{ac}G_{bc}^R G_{bd}^L-2\delta_{ad}\delta_{bd}G_{bc}^R G_{ac}^L-2\delta_{bc}\delta_{ac}G_{ad}^R G_{bd}^L-2\delta_{bc}\delta_{bd}G_{ad}^R G_{ac}^L\nn\\
&\quad &-2\delta_{bc}\delta_{ac}G_{ad}^R G_{bd}^L+2\delta_{ad}\delta_{bc}G_{bc}^R G_{ad}^L+2\delta_{ac}\delta_{bd}G_{ac}^R G_{bd}^L+4\delta_{cd}\delta_{cd}G_{ab}^R G_{ab}^L\nn\\
&=&-2\times 4G_{ab}^R G_{ab}^L+0+4\times 7G_{ab}^R G_{ab}^L\nn\\
&=&20G_{ab}^R G_{ab}^L,
\eeq
where identities $G_{ab}=-G_{ba}$, $\delta_{ab}G_{ab}=0$ and $\delta_{ab}\delta_{ab}=7$ are employed in passsing to the third equality. Then consider $(C\cdot G^R)(C\cdot G^L)$.
\beq
(C\cdot G^R)(C\cdot G^L)&=&4C_{dacm}G_{mb}^R C_{dacn}G_{nb}^L+4C_{dacm}G_{mb}^R C_{dabn}G_{nc}^L\nn\\
&=&4\times 24\delta_{mn}G_{mb}^R G_{nb}^L+4\times\left(-2C_{cmbn}+4(\delta_{cb}\delta_{mn}-\delta_{cn}\delta_{bm})\right)G_{mb}^R G_{nc}^L\nn\\
&=&96G_{ab}^R G_{ab}^L+8C_{abcd}G_{ab}^R G_{cd}^L+4G_{ab}^R G_{ab}^L-4\times 0\nn\\
&=&96G_{ab}^R G_{ab}^L+16G_{ab}^R G_{ab}^L+16G_{ab}^R G_{ab}^L\nn\\
&=&128G_{ab}^R G_{ab}^L,
\eeq
where identities $C_{mnab}C_{mncd}=-2C_{abcd}+4(\delta_{ac}\delta_{bd}-\delta_{ad}\delta_{bc})$ and $C_{abcm}C_{abcn}=24\delta_{mn}$ are employed in the second equality, and identities $C_{abcd}G_{cd}=2G_{ab}$ and $G_{ab}=-G_{ba}$ are employed in the fourth equality. Then consider $(\delta\cdot G^R)(C\cdot G^L)+(C\cdot G^R)(\delta\cdot G^L)$.
\beq
(\delta\cdot G^R)(C\cdot G^L)+(C\cdot G^R)(\delta\cdot G^L)&=&4C_{dacm}G_{mc}^R G_{ad}^L+4C_{bcam}G_{mb}^R G_{ac}^L\nn\\
&\quad &-4C_{dacm}G_{ma}^R G_{cd}^L-4C_{bcam}G_{bc}^R G_{ma}^L\nn\\
&=&4\times 4C_{abcd}G_{ab}^R G_{cd}^L\nn\\
&=&32G_{ab}^R G_{ab}^L,
\eeq
where identities $\delta_{ab}C_{abcd}=0$ and $C_{abcd}G_{cd}=2G_{ab}$ are employed. In summary, the OPE of $G_{ab}^R(z)G_{ab}^L(\bar{z})G_{cd}^R(w)G_{cd}^L(\bar{w})$ gives rise to
\beq
G_{ab}^R(z)G_{ab}^L(\bar{z})G_{cd}^R(w)G_{cd}^L(\bar{w})&\sim &\frac{1}{|z-w|^2}\frac{4}{9}\left(20+\frac{1}{16}\times 128+\frac{1}{4}\times 32\right)G_{ab}^R G_{ab}^L\nn\\
&=&\frac{1}{|z-w|^2}16G_{ab}^R G_{ab}^L.
\eeq

2.
\beq
T_{a}^R(z)T_{a}^L(\bar{z})T_{b}^R(w)T_{b}^L(\bar{w})&\sim &\frac{1}{|z-w|^2}\left(-G_{ab}^R+\frac{1}{\sqrt{3}}C_{abc}T_c^R\right)\left(-G_{ab}^L+\frac{1}{\sqrt{3}}C_{abd}T_d^L\right)\nn\\
&=&\frac{1}{|z-w|^2}\left( G_{ab}^R G_{ab}^L+2T_a^R T_a^L\right),
\eeq
where identities $C_{abc}G_{bc}=0$ and $C_{abc}C_{abd}=6\delta_{cd}$ are employed.

3.
\beq
V_{a}^R(z)V_{a}^L(\bar{z})V_{b}^R(w)V_{b}^L(\bar{w})&\sim &\frac{1}{|z-w|^2}\left(-G_{ab}^R-\frac{1}{\sqrt{3}}C_{abc}T_c^R\right)\left(-G_{ab}^L-\frac{1}{\sqrt{3}}C_{abd}T_d^L\right)\nn\\
&=&\frac{1}{|z-w|^2}\left( G_{ab}^R G_{ab}^L+2T_a^R T_a^L\right),
\eeq
where identities $C_{abc}G_{bc}=0$ and $C_{abc}C_{abd}=6\delta_{cd}$ are employed.

4.
\beq
G_{ab}^R(z)G_{ab}^L(\bar{z})T_{c}^R(w)T_{c}^L(\bar{w})&\sim &\frac{1}{|z-w|^2}\left(\frac{2}{3}\delta_{bc}T_a^R-\frac{2}{3}\delta_{ac}T_b^R+\frac{1}{3}C_{abcd}T_d^R\right)\nn\\
&\quad &\quad\quad\quad\quad~ \times\left(\frac{2}{3}\delta_{bc}T_a^L-\frac{2}{3}\delta_{ac}T_b^L+\frac{1}{3}C_{abce}T_e^L\right)\nn\\
&=&\frac{1}{|z-w|^2}\left(\frac{4}{9}(\delta_{bc}T_a^R-\delta_{ac}T_b^R)(\delta_{bc}T_a^L-\delta_{ac}T_b^L)\right.\nn\\
&\quad &\quad\quad\quad\quad~ +\left.\frac{1}{9}C_{abcd}C_{abce}T_d^R T_e^L\right)\nn\\
&=&\frac{1}{|z-w|^2}\frac{1}{9}(4\times 2\times 7T_a^R T_a^L-4\times 2T_a^R T_a^L+24\delta_{de}T_d^R T_e^L)\nn\\
&=&\frac{1}{|z-w|^2}8T_a^R T_a^L,
\eeq
where identities $\delta_{ab}\delta_{ab}=7$, $\delta_{ab}C_{abcd}=0$ and $C_{abcm}C_{abcn}=24\delta_{mn}$ are employed. Since both $T_a$ and $V_a$ forms $G_2$ vector representations, the result above can be directly generalized to the OPE below:
\beq
&&G_{ab}^R(z)G_{ab}^L(\bar{z})V_{c}^R(w)V_{c}^L(\bar{w})\sim \frac{1}{|z-w|^2}8V_a^R V_a^L,\\
&&G_{ab}^R(z)G_{ab}^L(\bar{z})\left(V_{c}^R(w)T_{c}^L(\bar{w})+T_{c}^R(w)V_{c}^L(\bar{w})\right)\sim \frac{1}{|z-w|^2}8\left(V_{a}^RT_{a}^L+T_{a}^RV_{a}^L\right).
\eeq

5.
\beq
T_{a}^R(z)T_{a}^L(\bar{z})V_{b}^R(w)V_{b}^L(\bar{w})&\sim &\frac{1}{|z-w|^2}\frac{1}{3}C_{abc}C_{abd}V_c^R V_d^L\nn\\
&=&\frac{1}{|z-w|^2}\frac{1}{3}\times 6\delta_{cd}V_c^R V_d^L\nn\\
&=&\frac{1}{|z-w|^2}2V_a^R V_a^L.
\eeq

6.
\beq
&\quad &\left(V_{a}^R(z)T_{a}^L(\bar{z})+T_{a}^R(z)V_{a}^L(\bar{z})\right)\left(V_{b}^R(w)T_{b}^L(\bar{w})+T_{b}^R(w)V_{b}^L(\bar{w})\right)\nn\\
&\sim &\frac{1}{|z-w|^2}2\left(\left(-G_{ab}^R+\frac{1}{\sqrt{3}}C_{abc}T_c^R\right)\left(-G_{ab}^L-\frac{1}{\sqrt{3}}C_{abd}T_d^L\right)\right.\nn\\
&\quad &\quad\quad\quad\quad\quad +\left.2\left(-\frac{1}{\sqrt{3}}C_{abc}V_c^R\right)\left(-\frac{1}{\sqrt{3}}C_{abd}V_d^L\right)\right)\nn\\
&=&\frac{1}{|z-w|^2}\left(2G_{ab}^R G_{ab}^L-4T_a^R T_a^L+4V_a^R V_a^L\right),
\eeq
where identities $C_{abc}G_{bc}=0$ and $C_{abc}C_{abd}=6\delta_{cd}$ are employed.

7.
\beq
&\quad &T_{a}^R(z)T_{a}^L(\bar{z})\left(V_{b}^R(w)T_{b}^L(\bar{w})+T_{b}^R(w)V_{b}^L(\bar{w})\right)\nn\\
&\sim &\frac{1}{|z-w|^2}\left(\left(-G_{ab}^R+\frac{1}{\sqrt{3}}C_{abc}T_c^R\right)\left(-\frac{1}{\sqrt{3}}C_{abd}V_d^L\right)\right.\nn\\
&\quad &\quad\quad\quad\quad~ +\left.\left(-\frac{1}{\sqrt{3}}C_{abc}V_c^R\right)\left(-G_{ab}^L-\frac{1}{\sqrt{3}}C_{abd}T_d^L\right)\right)\nn\\
&=&\frac{1}{|z-w|^2}\left(-2\left(V_{a}^RT_{a}^L+T_{a}^RV_{a}^L\right)\right),
\eeq
where identities $C_{abc}G_{bc}=0$ and $C_{abc}C_{abd}=6\delta_{cd}$ are employed.

8.
\beq
&\quad &V_{a}^R(z)V_{a}^L(\bar{z})\left(V_{b}^R(w)T_{b}^L(\bar{w})+T_{b}^R(w)V_{b}^L(\bar{w})\right)\nn\\
&\sim &\frac{1}{|z-w|^2}\left(\left(-G_{ab}^R-\frac{1}{\sqrt{3}}C_{abc}T_c^R\right)\left(-\frac{1}{\sqrt{3}}C_{abd}V_d^L\right)\right.\nn\\
&\quad &\quad\quad\quad\quad~-\left.\left(-\frac{1}{\sqrt{3}}C_{abc}V_c^R\right)\left(-G_{ab}^L-\frac{1}{\sqrt{3}}C_{abd}T_d^L\right)\right)\nn\\
&=&\frac{1}{|z-w|^2}2\left(V_{a}^RT_{a}^L+T_{a}^RV_{a}^L\right),
\eeq
where identities $C_{abc}G_{bc}=0$ and $C_{abc}C_{abd}=6\delta_{cd}$ are employed.

\begin{figure}[htbp]
\centering
\includegraphics[width=0.6\linewidth]{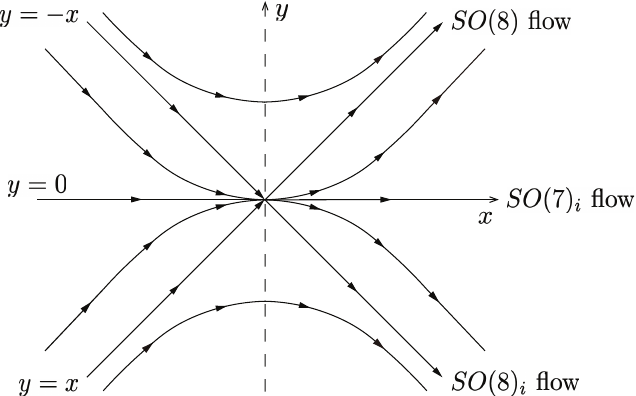}
\caption{RG flow on $SO(7)$ symmetric fixed planes.}\label{figSM1}
\end{figure}

Combine the results of OPE 1 $\sim$ 8. The OPE of interaction Hamiltonian density reads
\beq
&\quad &\mathcal{H}_\text{int}(z,\bar{z})\mathcal{H}_\text{int}(w,\bar{w})\nn\\
&\sim &\frac{1}{|z-w|^2}\Big((16g^2+t^2+y^2+w^2)\sum_{a,b}G_{ab}^R G_{ab}^L+(16gt+2t^2+2y^2-2w^2)\sum_{a}T_a^R T_a^L\nn\\
&\quad &+(16gy+4ty+2w^2)\sum_{a}V_a^R V_a^L+(16gw-4tw+4yw)\frac{1}{\sqrt{2}}\sum_{a}\left(T_a^R V_a^L+V_a^R T_a^L\right)\Big),\nn\\
\eeq
which gives rise to the RG equations:
\bea
&\frac{\text{d}g}{\text{d} l}=16g^2+t^2+y^2+w^2,\\
&\frac{\text{d}t}{\text{d} l}=16gt+2t^2+2y^2-2w^2,\\
&\frac{\text{d}y}{\text{d} l}=16gy+4ty+2w^2,\\
&\frac{\text{d}w}{\text{d} l}=16gw-4tw+4yw.
\eea
The RG equations can be solved analytically on $SO(7)$ symmetric fixed planes. On the $SO(7)_A$ symmetric fixed plane, namely $t=2g, w=0$, the RG equation have only two independent variables, which can be chosen as $y$ and $x=g+\frac{t}{2}$. The RG equations are reduced to
\beq
\frac{\text{d}y}{\text{d}x}=\frac{6xy}{5x^2+y^2},
\eeq
whose solution is $(x^2-y^2)^3+Cy^5=0$, $\forall C\in\mathbb{R}$.
This set of curves have asymptotic lines $y=x$, $y=-x$ and $y=0$, corresponding to the $SO(8)$ GN fixed flow $t=y=2g, w=0$, the $SO(8)_A$ GN fixed flow $t=-y=2g, w=0$, and the $SO(7)_A$ GN fixed flow $t=2g, y=w=0$, respectively. The plot of these solution curves are shown in figure~\ref{figSM1}, which looks the same as figure~\ref{fig:plane} in the main text. According to the triality, the RG flows on the other two $SO(7)$ symmetric fixed planes are the same.

The RG equations can be also investigated in terms of $g,~t_A,~t_B,~t_M$ by employing the relation between $T_a'$, $V_a'$, $T_a''$, $V_a''$ and $T_a$, $V_a$. The OPE reads
\beq
&\quad &\mathcal{H}_\text{int}(z,\bar{z})\mathcal{H}_\text{int}(w,\bar{w})\nn\\
&\sim &\frac{1}{|z-w|^2}\Bigg(\Big(16g^2+t_A^2+t_B^2+t_M^2+\frac{1}{2}t_At_B+\frac{1}{2}t_Bt_M+\frac{1}{2}t_Mt_A\Big)\sum_{a,b}G_{ab}^R G_{ab}^L\nn\\
&\quad & \quad\quad\quad\quad~ +\left(16gt_A+2t_A^2+4t_Bt_M\right)\sum_{a}T_a^R T_a^L+\left. \left(16gt_B+2t_B^2+4t_Mt_A\right)\sum_{a}T_a^{\prime R} T_a^{\prime L}\right.\nn\\
&\quad &\quad\quad\quad\quad~ +\left(16gt_M+2t_M^2+4t_At_B\right)\sum_{a}T_a^{\prime\prime R} T_a^{\prime\prime L}\Bigg),
\eeq
which gives rise to the RG equations in the main text:
\bea
&\frac{\text{d}g}{\text{d} l}=16g^2+\sum_i t_i^2+\frac{1}{4}\sum_{i\neq j}t_
it_j,\\
&\frac{\text{d}t_i}{\text{d} l}=16gt_i+2t_i^2+4\frac{t_A t_B t_M}{t_i},\label{eqRGG}
\eea
where $i,j,k$ range in $A,B,M$. The RG flows on $SO(7)$ fixed planes are the same as figure~\ref{figSM1}.

Except for fixed flows and fixed planes, the RG equations (\ref{eqRGG}) also have fixed bodies, i.e. three dimensional bodies in the parameter space in which RG flows keep inside evolution. From (\ref{eqRGG}), it is not hard to derive that
\beq
&&\frac{\text{d}}{\text{d} l}\Big(\sum_it_i\Big) =16g\Big(\sum_it_i\Big)+2\Big(\sum_it_i\Big)^2\\
&&\frac{\text{d}}{\text{d} l}(t_i-t_j)=\Big(16g+2t_i+2t_j-4\sum_k\epsilon_{ijk} t_k\Big)(t_i-t_j).
\eeq
Therefore, both $\sum_it_i=0$ and $t_i=t_j$ are solutions to the RG equations (\ref{eqRGG}), and hence define four fixed bodies in the parameter space: $t_A+t_B+t_M=0$, $t_A=t_B$, $t_B=t_M$, and $t_M=t_A$. The intersections of these fixed bodies will give rise to some fixed planes.

\section{Bosonization of the interaction on $SO(7)$ symmetric fixed planes}\label{app:C}

For simplicity, we use the interaction Hamiltonian density on $SO(7)_M$ symmetric fixed plane which is directly written in terms of Majorana operators defined in (\ref{eq:31}). Hamiltonian densities on the other two $SO(7)$ symmetric fixed planes can be similarly solved via triality relation. 
\beq
\mathcal{H}_{M}=g\sum_{1\le m,n\le 7}(i\chi_m^R\chi_n^R)(i\chi_m^L\chi_n^L)+3(2g-t_M)\sum_{1\le m\le 7}(i\chi_0^R\chi_m^R)(i\chi_0^L\chi_m^L).
\eeq
It can be rearranged into a Gross-Neveu form (up to a constant)
\beq
\mathcal{H}_{M}=-g\Bigg(\sum_{1\le m\le 7}i\chi_m^R\chi_m^L+\left(3-\frac{3t_M}{2g}\right)i\chi_0^R\chi_0^L\Bigg)^2.
\eeq
Therefore, when $t_M=2g$, i.e. on the $SO(7)_M$ GN fixed flow, $\chi_0$ is decoupled in the interaction and becomes massless, corresponding to an Ising phase transition with central charge $c=\frac{1}{2}$.

According to a standard bosonization dictionary
\beq
i\chi_4^R\chi_4^L+i\chi_0^R\chi_0^L 
\sim \frac{1}{\pi a}\cos \sqrt{4\pi}\phi_\frac{3}{2},~~~i\chi_4^R\chi_4^L-i\chi_0^R\chi_0^L 
\sim \frac{1}{\pi a}\cos \sqrt{4\pi}\theta_\frac{3}{2},
\eeq
the Hamiltonian density above can be bosonized as
\beq
\mathcal{H}_{M}=&-&\frac{g}{(\pi a)^2}\Bigg(\sum_{\sigma\neq\frac{3}{2}}\cos \sqrt{4\pi}\phi_\sigma+\left(2-\frac{3t_M}{4g}\right)\cos \sqrt{4\pi}\phi_\frac{3}{2}+\left(-1+\frac{3t_M}{4g}\right)\cos \sqrt{4\pi}\theta_\frac{3}{2}\Bigg)^2.\nn\\
\eeq
where in the second equality the derivative terms $-\frac{2}{\pi}(\partial_\mu\phi)^2=(:\cos\sqrt{4\pi}\phi :)^2$ are dropped. Consider $g>0$ case where the interactions are always relevant. For all $t_M$, the local minima of this bosonized Hamiltonian have $\phi_\sigma$, $\sigma\neq\frac{3}{2}$ pinned. For $\sigma=\frac{3}{2}$, the local minima have $\phi_\frac{3}{2}$ pinned when $t_M<2g$, and $\theta_\frac{3}{2}$ pinned when $t_M>2g$. This also suggests that $t_M=2g$ is an Ising transition point with central charge $c=\frac{1}{2}$. When $t_M=\frac{4}{3}g$ or $\frac{8}{3}g$, the bosonized Hamiltonian has $SO(8)$ symmetries\, which is consistent with $SO(8)$ and $SO(8)_M$ fixed flows obtained from RG equations, respectively.

\section{RG for the lattice model}\label{app:D}

We numerically integrate the RG equations on the lattice model phase diagram $P_\mathrm{Lat}:~t_A+t_B=0,~t_M=-4g$. Since $u(v)=0$ on the $v(u)$-axis, lattice model couplings $u$ and $v$ are related to continuum model parameters as $u=C_0(t_A-kg)$ and $v=C_0(t_B-kg)$. $C_0>0$ is an undetermined constant related to the details of coarse-graining and cutoff, which is irrelevant to our discussions. The initial values for RG equations are chosen as
\beq
g(0)=r_0\cos\theta,~ t_A(0)=-4r_0\sin\theta,~ t_B(0)=4r_0\sin\theta,~ t_M(0)=-4r_0\cos\theta,
\eeq
where $\theta$ varies from 0 to $2\pi$ and $r_0>0$. Thus, the initial values of $u$ and $v$ are
\beq
u(0)=R_0\cos\left(\theta+\frac{\pi}{4}\right),~v(0)=R_0\sin\left(\theta+\frac{\pi}{4}\right),
\eeq 
where $R_0=2\sqrt{2}C_0r_0>0$. Therefore, $u=v>0$ axis, positive $v$-axis, negative $u$-axis, $u=v<0$ axis, negative $v$-axis and positive $u$-axis are corresponding to $\theta=0$, $\frac{1}{4}\pi$, $\frac{3}{4}\pi$, $\pi$, $\frac{5}{4}\pi$ and $\frac{7}{4}\pi$. In numerics we find a critical $\theta_c=(0.705\pm 0.002)\pi$ for all $r_0>0$, such that when $0<\theta<\theta_c$, the RG flows are controlled by the $SO(8)_B$ GN fixed flow, suggesting the gapped $SO(8)_B$ GN phase; when $\theta_c<\theta<\pi$ and $\pi<\theta<2\pi-\theta_c$, the RG flows are controlled by the $SO(8)$ GN fixed flow, suggesting the $SO(8)$ GN phase; and when $2\pi-\theta_c<\theta<2\pi$, the RG flows are controlled by the $SO(8)_A$ GN fixed point, suggesting the $SO(8)_A$ GN phase. In numerics, we locate $\theta_c$ by the sign change of $t_A$ and $t_M$ around it which suggests a phase transition controlled by the $SO(7)_B$ GN fixed flow with central charge $c=\frac{1}{2}$. The situation of $\theta=2\pi-\theta_c$ is similar, where the phase transition is controlled by the $SO(7)_A$ GN fixed flow. When $\theta=0$ and $\theta=\pi$, the RG flows are controlled by the $G_2$ TCI and the $SO(7)_M$ GN fixed flows, suggesting phase transitions with central charges $c=\frac{6}{5}$ and $c=\frac{1}{2}$, respectively. Note that
\beq
\frac{v(0)}{u(0)}=\tan\left(\theta_c+\frac{\pi}{4}\right)\approx-\frac{1}{7},
\eeq
with a deviation in $\theta_c$ (with respect to $2\pi$) less than $0.2\%$. This suggests that the equation of the $SO(7)_B$ phase transition line is consistent with the predicted location $u=-7v$ based on a mean-field analysis. The acceptably small deviation of $0.2\%$ arises from the perturbative nature of the one-loop RG method. By a similar argument, the $SO(7)_A$ transition line is $v\approx-7u$.

\end{document}